# MEDICAL MASK WITH PLASMA STERILIZING LAYER

Andrey Y. Starikovsky[1*], Dinara R. Usmanova[2]

[1]Mechanical and Aerospace Department, Princeton University, Princeton, USA;
[2]Department of Systems Biology, Columbia University, New York, USA
*astariko@princeton.edu

In this brief report we propose a new design of a medical mask with a plasma layer, which provides both additional air filtration from microdrops, bacteria and viruses due to the electrostatic effect and self-disinfecting of surfaces by a pulsed barrier discharge. The key features of the mask are the mutual arrangement of the layers, the direction of air flows and the synchronization of the discharge with respiration, which ensures the safe wearing of the mask and high degree of protection against pathogenic microorganisms.

**Introduction**

During the current spread of the pandemic of COVID-19 [1] the problem of individual protection against airborne transmission gets very pressing. Airborne transmission happens when droplets containing viruses or bacteria get on mucous membranes [2]. There are two main avenues for the spread of COVID-19. One way involves contacts with contaminated surfaces and subsequent touching mouth, nose or eyes. However, the majority of transmissions is believed to happen by person-to-person spread through respiratory droplets [3]. Such droplets are produced when infected person coughs or sneezes and can spread up to two meters and stay in the air several hours [4].

Usage of special respirators which apply an external stream of purified air is an efficient method of protection against airborne transmission of the virus. Usually such devices are complicated, expensive, and require periodical filter replacement. Alternatively, simple masks don't provide complete protection to the wearer from viruses because of loose-fit or/and because the size of pores is larger than the size of the virus [5]. Even the usage of such masks for protection of others nearby from potential contamination is limited to a short period after which the mask itself becomes a source of contamination [6].

In this brief communication we propose a modification of the existing masks which will provide an additional filtration of the air and deactivation of microbes by low-temperature plasma.

**Medical mask design**

The standard N-95 or N-99 mask is comprised of several layers (figure 1) [7]. Consequent layers are made from different filtering materials and remove droplets and microparticles of smaller and smaller sizes. Moreover, middle layer consists of activated carbon which can absorb harmful gases, for example, $NO_x$, CO, $CH_x$, $O_3$. Such masks remove 95% (N-95) and 99% (N-99) of particles larger than 0.3 microns from the air stream [7].

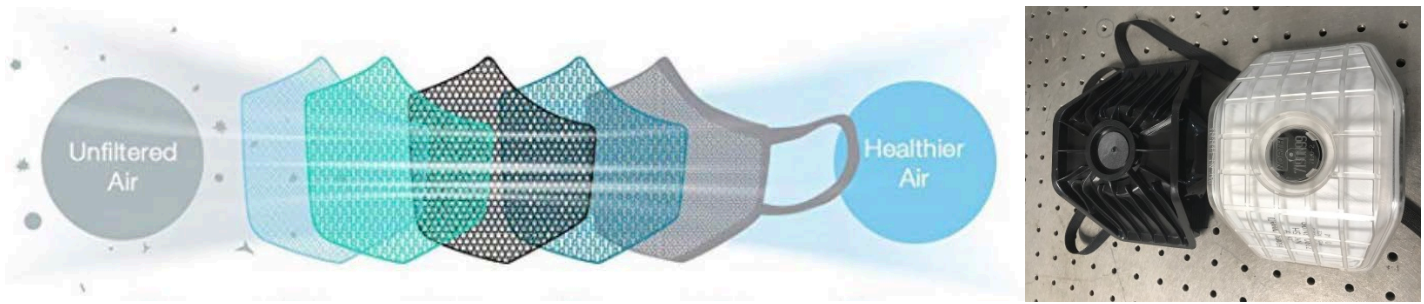

Figure 1. The standard sequence of layers in N-95/N-99 masks (left); N-99 half-face respirator (right).

Such design of the mask substantially reduces the probability of contamination. However, it can't completely protect from the virus reaching the organism (the size of SARS-CoV-2 virus particle is ~ 100 nm [8]). Moreover, after prolonged wearing all layers of the mask accumulate various pathogens and themselves become the source of infection [6]. We propose to solve both problems by adding two extra layers to the standard mask: i) plasma layer, formed by dielectric barrier discharge, and ii) the layer which protects the skin from direct exposure to the plasma.

Figure 2 shows two new layers that are proposed to be added to the standard N-95/99 respirator. The first additional layer (farthest from the skin) is a standard dielectric barrier discharge (DBD) formed by wire electrodes surrounded by a dielectric layer. The polarity of the electrodes alternates, which leads to the formation of a nonequilibrium low-temperature plasma between the walls of individual dielectric tubes. It is known [9-13] that such plasma is extremely active antimicrobial media.

The main mechanisms of plasma impact on the microorganisms are 1) heat; 2) ultraviolet radiation of the discharge; 3) nitrogen oxides ($NO_x$); 4) active oxidizing agents formed in plasma ($O_3$, $H_2O_2$); 5) positive and negative ions. In our system we expect the major disinfection effect to originate from (4) and (5). In the case of nonequilibrium air plasma, gas heating is negligible and does not exceed tens of degrees. This increase is sufficient for the destruction of biomolecules directly in the region of the discharge, but even at a small distance from it the thermal action of the plasma on bacteria and viruses can be neglected. The major part of UV radiation of air DBD plasma is concentrated in the bands of the second positive system of nitrogen (with the strongest transition at 337.1 nm). Such quanta usually don't break bonds in biomolecules and are not an effective means of disinfection (unlike, for example, mercury radiation at 253.7 nm, which leads to irreversible DNA/RNA damage and stops the growth of bacteria and viruses [9,10]). Plasma nitrogen oxides are generally regarded as stimulants of the natural response of the immune system. The main mechanism of action of oxidizing agents ($O_3$, $H_2O_2$) on viruses is the destruction of their receptors. The loss of receptors disables virus ability of attacking human cell, which makes its reproduction impossible. Positive and negative ions act on viruses in the similar way as molecular oxidizing agents, often enhancing the effect of the latter [11].

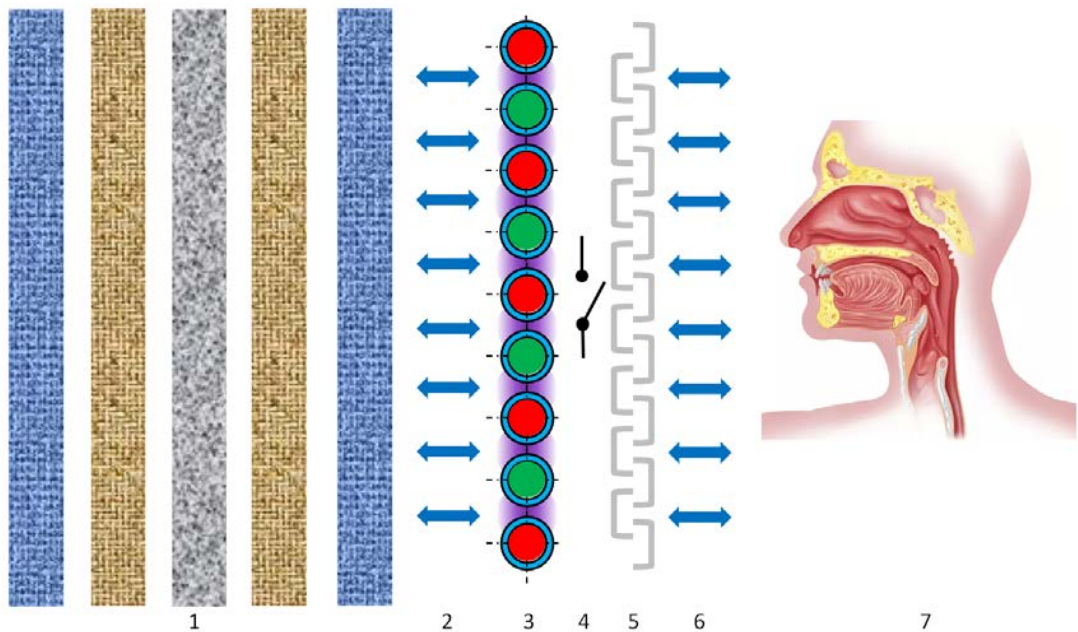

Figure 2. The scheme of the layers of the modified respirator mask. 1 - standard five-layer design; 2 and 6 - bidirectional air flow through the filters of the mask; 3 - electrode system. Green and red dots represent electrodes of different alternating polarity mounted in dielectric capillaries (blue). The discharge develops in the gap between the dielectric capillaries, where the entire air flow passes; 4 - switch, sensitive to the direction of the air flow. It turns on the discharge during the expiration, and disables it during the inspiration; 5 - a dielectric layer providing the necessary electrical gap between the skin and the electrode system, and blocking the ultraviolet radiation of the discharge; 7 – mask wearer.

In addition to mechanisms discussed above, another effect becomes important – the electrostatic precipitation. It happens because microdrops and single microorganisms in the air usually accumulate a noticeable negative charge due to the high mobility of free electrons in the air. Negatively charged drops and microparticles could be carried away from the flow to the surface by the electric field. Thus, a small residual positive charge on the surface of the DBD electrodes mechanically removes microparticles from the stream. Moreover, it concentrates them in the discharge zone, which then is completely disinfected by the action of plasma.

The problems associated with the use of plasma in medical equipment are well known [9]. Firstly, there is a potentially dangerous high voltage on the electrodes of the system. However, the danger to health is not voltage, but current. In the case of our design, the peak current (~ 2 mA) generated by the power supply is at least an order of magnitude less than the minimal current able to cause any muscle reaction [14]. Such currents cannot lead to any dangerous consequences even in the case of direct contact of the high-voltage electrode with the skin. In addition, all electrodes are surrounded by the dielectric layer, which makes direct contact with high voltage electrodes impossible. The alternation of electrodes of different polarity (fig. 2) leads to a sharp decrease in the electric field when moving away from the plane of the electrode system. In fact, already at a distance of the diameter of the dielectric tube from the electrode, the electric field is close to zero.

Radiation of the second positive system of nitrogen in the range 300-350 nm, which is typical for discharges in air, can cause skin sunburn in case of prolonged exposure [15]. As already noted, radiation of this range cannot lead to DNA/RNA damage and doesn't cause a skin cancer. However, substantial doses of such radiation are not healthy and should be blocked. To block the UV radiation of the discharge, the dielectric screen should be made from an opaque UV-blocking material with overlapping elements (fig. 2).

Another problem of air disinfection using air plasma is the high carcinogenicity of the generated oxidants - in particular, ozone $O_3$. To solve this issue, we propose the synchronization of the electric discharge with the respiration. During the inspiration phase, the discharge is completely turned off, and the production of ozone and other active components is completely stopped. The mask works as a standard five-layer filter with an additional electrostatic filter at the final stage. This electrostatic filter works by accumulating residual positive charge on the dielectric surfaces of the electrodes (fig.2, element 3) and of the dielectric screen (fig.2, element 5), and does not create dangerous components in the inhaled air stream.

On the contrary, during the expiration phase the discharge is switched on. It disinfects the surface of the electrodes, exhaled air, and filters of the standard set. This eliminates the contact of ozone with the skin of the mask wearer. Moreover, chemical reactions with ozone in the outer layers of the mask prevent a significant increase of ozone concentration in surrounding air.

To power the discharge we propose to use a bipolar periodic voltage applied to the adjacent electrodes of the plasma system (fig. 2). The shape of the pulses can be arbitrary, but to reduce the energy consumption and therefore, to increase the operating time of the device from one set of batteries, it is easier to use rectangular pulses created by a symmetric multivibrator with a transformer. The amplitude of the pulses should be selected based on the characteristic distance between the electrodes. With a typical diameter of the dielectric $d$ = 1 mm and the open area ~10% of the layer (gap between electrodes is $\delta$ ~ 0.1 mm), the peak-to-peak voltage at the electrodes should be ~ 3 kV (1.5 kV amplitude). With this voltage, a high electric field appears in the gap between the dielectric surfaces, a discharge develops in the gap, and the plasma layer spreads over the surface of the dielectrics, ensuring their complete sterilization.

The pulse frequency should be high enough to ensure that at least several pulses are generated during the exhalation phase. Note that the inspiratory phase is not important for us, since at this phase the discharge is completely turned off. During the exhalation phase, the typical duration of which is about 1-2 seconds, it is possible to achieve the development of tens of discharges even with pulse frequency of ~ 100 Hz. Ozone, ions, electronically excited plasma particles and UV radiation generated by DBD will actively sterilize the inner layers of the standard mask filters and the surface of the electrodes, which play the role of an electrostatic filter in our system (fig. 2, elements 1 and 3). Note that neither the UV radiation nor the active components generated in the discharge could reach either the skin or the lungs of the mask wearer, which guarantees the safety of using the mask.

**Experimental validation**

For experimental verification of the design shown in figure 2 we assembled an electrode system of a size comparable to the real N-99 mask (fig. 1). Wire electrodes were inserted into the quartz capillaries 6 cm long, external diameter 0.8 mm and internal diameter 0.4 mm. The electrodes were alternately connected to the positive and negative terminals of the pulsed high-voltage power supply. An assembly of 6×6 cm² was formed by the electrodes maintaining a gap between the individual electrodes of $\delta$ ~ 0.1 mm. The size of the gap ensured a low aerodynamic drag of the assembly and good uniformity of the flow (fig. 3).

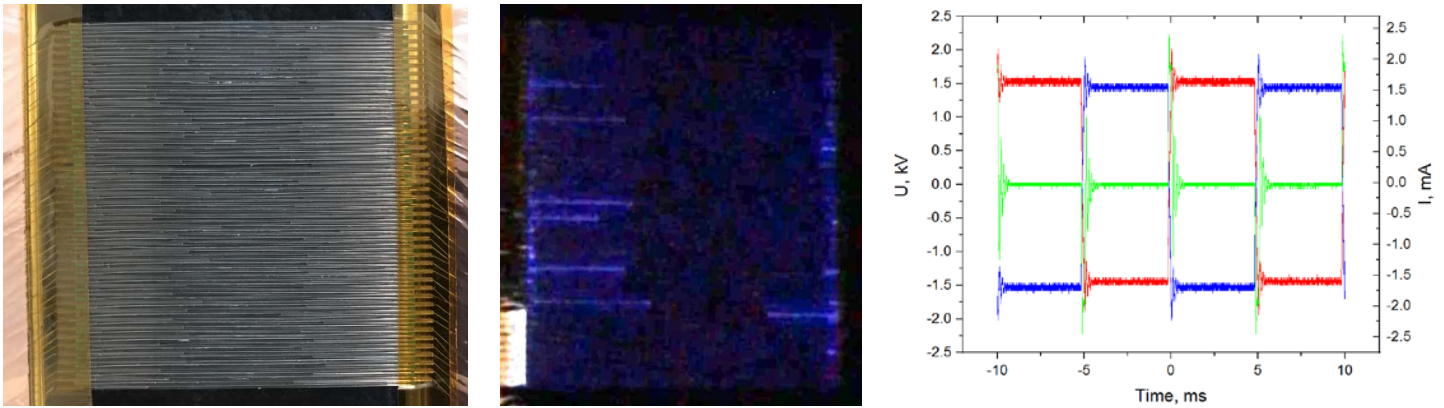

Figure 3. The design of the electrodes of the barrier discharge layer (left); plasma emission at maximum test voltage of 4 kV (center); voltage and current through the discharge device in the design regime (right). Red and blue lines represent voltage on the adjacent electrodes, green line - total current.

The system demonstrated a good uniformity of the generated plasma in a wide range of voltages and pulse frequencies. The system was tested to a maximum amplitude voltage of ±4 kV (fig. 3, center). The stable discharge emerges at the electrode voltage above $U_0$ = ±1.5 kV and frequency $f$ = 100 Hz. At the moment of polarity switching, the peak current through the device reached $I_{peak}$ = 2 mA and the duration of the current pulse was about $\Delta t$ = 100 µs (fig. 3, right). Thus the total power of the high voltage discharge was $P \sim U_0 I_{peak} \Delta t (2f)$ = 0.12 W. It means that a standard 6 V battery with a capacity of 4,500 mAh could be enough for more than a week of continuous operation of the device. An increase in the size of the discharge section will lead to an increase in energy consumption and therefore to the reduction of the operation time. For example, covering the mask N-99 with the filter area $S$ ~ 10×20 cm² (fig. 1) will reduce the time of continuous operation of the plasma sterilization layer to 1-2 days.

An average specific current of dielectric barrier discharge (fig. 3) at applied voltage $U$ = 1.5 kV reached $i = I_{peak} \Delta t (2f)/S$ = 1.1 µA/cm². Work [11] has demonstrated that 6-log bacteria inactivation by plasma at sufficiently high electric field (several times above the breakdown threshold) requires a dose of $q$ ~ 500 µC/cm². It means that the plasma layer in a medical mask operating even at minimal discharge power will provide a 6-log reduction of pathogens every $\tau = q/i$ ~ 10 minutes.

The usage of quartz capillary tubes as dielectric layers for barrier discharge is not the best choice for mass production. A good replacement for this design is thin high-voltage wires in PFTE insulation (for example, type MIL-W-16878/4). Such wires with an outer diameter of 0.76 mm can operate in continuous regime up to 600 V. Above this voltage a corona discharge could appear around the external dielectric of the wire. It is not acceptable for common electronic devices but it is namely the regime required for the DBD plasma generation. The thickness of the PFTE dielectric layer of such a wire is 0.25 mm (breakdown voltage of 6 kV), which ensures its durability as an electrode for dielectric barrier discharge up to voltages of 1.5-2 kV (3-4 kV peak-to-peak) with a sufficient safety margin. PFTE insulation is able to withstand the temperature increase up to 200° C without damage, which makes it possible to sterilize the device at high temperature, if required.

**Conclusions**

A simple design of a self-disinfecting medical mask based on a pulse-periodic barrier discharge is proposed. The plasma created by such a discharge has a high uniformity and a high efficiency in the production of biologically active components in the air. Therefore, proposed modifications increase the service life and the effectiveness of the standard medical masks and significantly improve the degree of protection for medical personnel who are constantly in contact with infected patients.